\documentclass[lnicst,sechang,a4paper]{svmultln}
\usepackage{makeidx, graphicx, amssymb}  
\begin{document}
\mainmatter              
\title{SIRS dynamics on random networks:\\ 
simulations and analytical models}
\titlerunning{SIRS dynamics on random networks}  
%
\author{Ganna Rozhnova \and Ana Nunes}
\authorrunning{Ganna Rozhnova et al.}   
%
\tocauthor{Ganna Rozhnova, Ana Nunes}

\institute{Centro de F{\'\i}sica Te{\'o}rica e Computacional and
Departamento de F{\'\i}sica,\\
Faculdade de Ci{\^e}ncias da Universidade de
Lisboa,\\
P-1649-003 Lisboa Codex, Portugal\\
\email{a_rozhnova@cii.fc.ul.pt}}

\maketitle              

\begin{abstract}        
The standard pair approximation equations (PA) for the Susceptible-Infective-Recovered-Susceptible (SIRS) model of infection spread on a network of homogeneous degree $k$ predict a thin phase of sustained oscillations for parameter values that correspond to diseases that confer long lasting immunity. Here we present a study of the dependence of this oscillatory phase on the parameter $k$ and of its relevance to understand the behaviour of simulations on networks. For $k=4$, we compare the phase diagram of the PA model with the results of simulations on regular random graphs (RRG) of the same degree. We show that for parameter values in the oscillatory phase, and even for large system sizes, the simulations either die out or exhibit damped oscillations, depending on the initial conditions. This failure of the standard PA model to capture the qualitative behaviour of the simulations on large RRGs is currently being investigated.
\end{abstract}
A number of approaches has been used to study the spreading dynamics of an infectious disease. A common paradigm, emerging from a simple deterministic framework, is to assume that populations are not spatially distributed so that individuals mix perfectly and contact each other with equal probability. Thus in the limit of infinite populations, the time evolution of the disease is described in terms of the densities of infectives and susceptibles as a function of time, and  governed by a system of coupled ordinary differential equations which can be deduced from the law of mass action \cite{murray}. Another approach is to use stochastic dynamics on a lattice (or more general graphs) where the variables at each node represent the state of an individual. 
The effects of spatial correlations that mass action models disregard play an important
role in the behaviour of infection dynamics on graphs, and therefore also in real populations
\cite{contact_rev}. The ordinary pair approximation (PA) as well as various improvements to include higher order correlations have been proposed in the context of ecological and epidemiological deterministic models \cite{ref_PA}. In {\cite{lebowitz}}, the performance of the PA in the description of the steady states and the dynamics of the Susceptible-Infective-Recovered-Susceptible (SIRS) model on the hypercubic lattice was analyzed in detail.

In this study, we consider the dynamics of the same epidemic model on a random network of homogeneous degree $k$ and $N$ nodes, a regular random graph of degree $k$ (RRG-$k$). Each node can be occupied by an individual in susceptible ($S$), infected ($I$), or recovered ($R$) state. Infected individuals recover at rate $\delta$, recovered individuals lose immunity at rate $\gamma$, and infection of the susceptible node occurs at infection rate $\lambda$ multiplied by the number of its infected nearest neighbours $n$, $n \in \{0,1,\ldots,k\}$:
\begin{eqnarray}
\label{sirs_rates} 
I\stackrel{\delta}{\rightarrow}R \ , \nonumber \\
R\stackrel{\gamma}{\rightarrow}S  \ , \\
S\stackrel{\lambda n}{\rightarrow} I \ . \nonumber
\end{eqnarray}

In the infinite population limit, with the assumptions of spatial homogeneity and uncorrelated pairs, the system is described by the deterministic equations of the standard or
uncorrelated PA \cite{lebowitz}: 
\begin{eqnarray}
\frac{d\:s}{d\:t}&=&\gamma \ (1-i-s)-k\lambda \ si \ , \nonumber\\
\frac{d\:i}{d\:t}&=&k\lambda \ si-\delta \ i \ , \nonumber\\
\label{paeq} \frac{d\:si}{d\:t}&=&\gamma \ ri-(\lambda+\delta) \ si + \frac{(k-1)\lambda \ si}{s}(s-sr-2si) \ , \\
\frac{d\:sr}{d\:t}&=&\delta \ si+\gamma \ (1-s-i-ri-2sr)-\frac{(k-1)\lambda \ si \ sr}{s} \ , \nonumber\\
\frac{d\:ri}{d\:t}&=&\delta \ (i-si)-(\gamma+2\delta) \ ri+\frac{(k-1)\lambda \ si \ sr}{s} \ . \nonumber
\end{eqnarray}
In the above equations the variables $s$, $i$ stand for the probability that a randomly chosen node is in state $S$, $I$, and the variables $si$, $sr$, $ri$ stand for the probability that a randomly chosen pair of nearest neighbour nodes is an $SI$, $SR$, $RI$ pair. 
As expected, neglecting the pair correlations and setting the pair state probabilities equal to the product of the node state probabilities these equations reduce to the classic equations of the randomly mixed  SIRS model.

\begin{figure}
\centering
\includegraphics[width=\textwidth]{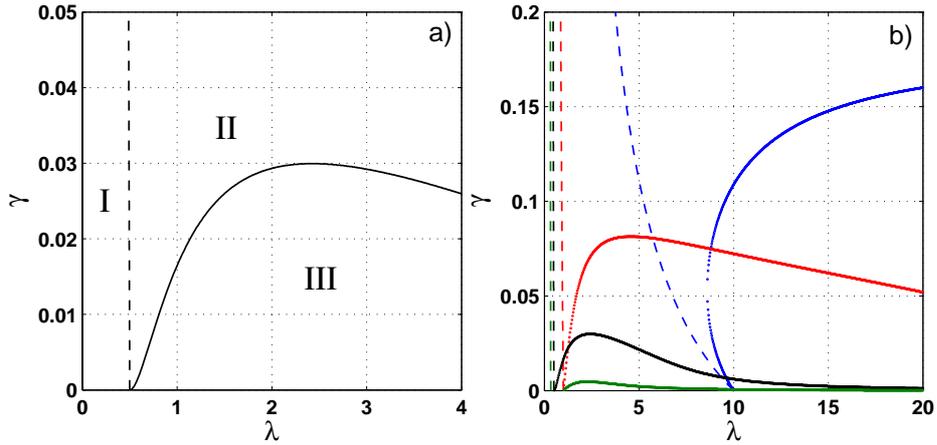}
\caption{a) Phase
diagram in the $(\lambda,\gamma)$ plane for the PA SIRS model, where
region I corresponds to susceptible-absorbing states and
region II corresponds to active states with nonzero infective
densities. The critical line between regions I and II is the black
dashed line. The second critical curve (black solid line) bounds a region with
limit cycle solutions (region III). Parameters: $\delta=1$, $k=4$. b) Phase diagram of the 
PA SIRS model for $\delta=1$ and $k=2.1$ (blue), $k=3$ (red), $k=4$ (black), $k=5$ (green). Dashed (dotted) lines correspond to the transcritical (supercritical Hopf) bifurcation curves.}
\label{bif_diagrams}
\end{figure}

The phase diagram of the PA SIRS model (\ref{paeq}) for $k=4$ is plotted in Fig. \ref{bif_diagrams}a).
We have set the time scale so that $\delta =1$.
Region I represents susceptible-absorbing states and region II corresponds
to active states that can be asymptotically stable nodes or asymptotically stable foci.
The critical line separating regions I and II corresponds to the transcritical bifurcation curve that is given by $\lambda_{c}(\gamma)=(\gamma +1)/(3\gamma+2)$ (black dashed line).
In addition, for small values of $\gamma $ we find a new phase boundary (black solid line), that corresponds to a supercritical Hopf bifurcation of the nontrivial equilibrium and has been missed in previous studies of this model {\cite{lebowitz}}. This boundary separates the active phase with constant densities from an active phase with oscillatory behavior, that is stable at low $\gamma $.  

In the thin phase of region III, the PA model predicts sustained oscillations in the thermodynamic limit. We have performed a systematic study of the dependence of this oscillatory phase on the parameter $k$ and of its relevance to understand the behaviour of simulations on networks. 

The phase diagram of the PA model for $\delta=1$ and several values of $k$ in the range $k>2$ is shown in 
Fig. \ref{bif_diagrams}b). The critical lines separating the absorbing and the active
phases (dashed lines) are given by $\lambda_{c}(\gamma)=(\gamma +1)/((k-1)\gamma+k-2)$. 
Within the active phase, the dotted lines are numerical plots of Hopf bifurcation curves. 
The oscillatory phase is large for $k \gtrsim 2$ and it gets thinner as $k$ increases,
but it persists for the whole range of $2< k \lesssim 6$. A similar phase diagram, with a Hopf bifurcation critical line bounding an oscillatory phase, was reported in other studies of related models \cite{rand_morris_jerome}, where SIR dynamics with different mechanisms of replenishment of susceptibles is modelled at the level of pairs with the standard or another closure approximation. These different models all exhibit an oscillatory phase in the regime of slow driving  through introduction of new susceptible individuals (small $\gamma $ in the present case). This suggests that this oscillatory phase may be related with the phenomenon of recurrent epidemics in infectious diseases that confer permanent or long lasting immunity.    

We have compared the behaviour of the PA SIRS model (\ref{paeq}) for $k=4$ with the results of stochastic simulations on RRG-$4$ for several system sizes. In the stochastic simulations, the system was set in a random initial condition with given node and pair densities and an efficient algorithm for stochastic processes in spatially structured systems 
based on Gillespie's method \cite{gillespie} was used to update
the states of the nodes according to the processes of infection, recovery and immunity waning (\ref{sirs_rates}). 
For each set of parameter values and initial conditions, the simulations were averaged
over $10^3$ realizations of the RRG-$4$ graph. 

\begin{figure}
\centering
\includegraphics[width=\textwidth]{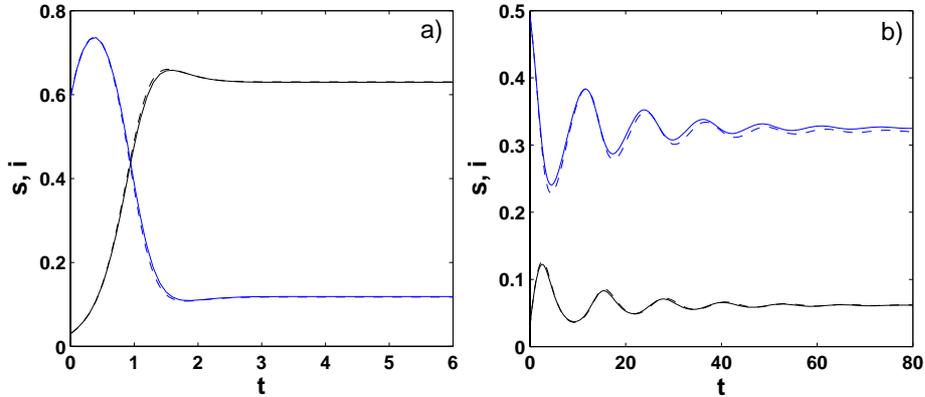}
\caption{For $\delta=1$, $k=4$, comparison of the solutions of the PA deterministic model (dashed lines) with the results of stochastic simulations (solid lines) on a RRG-$4$ with $N=10^6$ for parameter values in region II. Susceptible (infective) densities are plotted in blue (black). Parameters: a) $\gamma=2.5$, $\lambda=2.5$; b) $\gamma=0.1$, $\lambda=2.5$.}
\label{timeseries}
\end{figure}

\begin{figure}
\centering
\includegraphics[width=\textwidth]{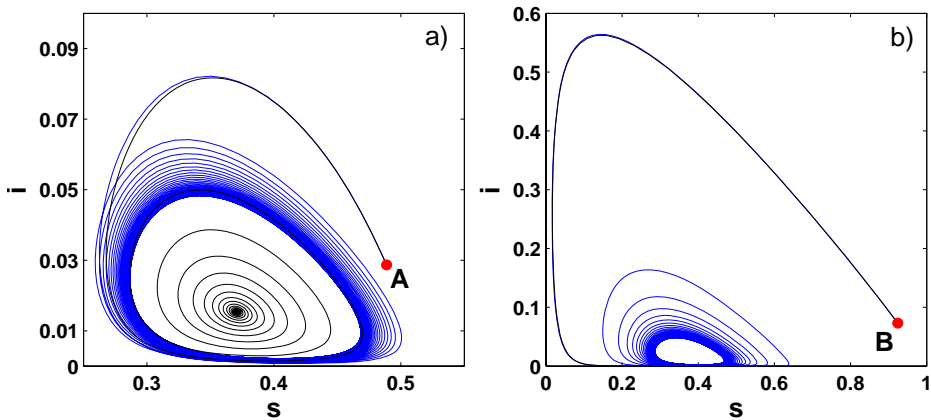}
\caption{For $\delta=1$, $k=4$, comparison of the solutions of the PA deterministic model (blue solid line) with the results of stochastic simulations (black solid line) on a RRG-$4$ with $N=10^6$ for parameter values in region III, $\gamma=0.025$, $\lambda=2.5$, and two sets of initial conditions $A$ and $B$ marked by red dots in the $(s,i)$ plane. Initial conditions: a) $s\approx0.4889$, $i\approx0.0287$, $si\approx0.0104$, $sr\approx0.2369$, $ri\approx0.0129$; b) $s\approx0.9240$, $i\approx0.0731$, $si\approx0.0558$, $sr\approx0.0024$, $ri\approx0.0005$.}
\label{cycles}
\end{figure}

The results of stochastic simulations for $N=10^6$ and solutions of the PA SIRS equations (\ref{paeq}) are shown in Figs. \ref{timeseries} and \ref{cycles}. The susceptible (blue lines) and the infective (black lines) densities are shown in  Fig. \ref{timeseries} for 
two sets of parameter values: $\gamma=2.5$, $\lambda=2.5$ (Fig. \ref{timeseries}a)) and $\gamma=0.1$, $\lambda=2.5$ (Fig. \ref{timeseries}b)). 
The numerical solutions of the PA SIRS equations are plotted in dashed lines, and the results of the simulations in
solid lines.
For parameter values well within region II of the phase diagram as in Fig. \ref{timeseries}a)
there is excellent agreement between the solutions of the PA SIRS model for the same initial densities and
the results of the stochastic simulations, both for the transient behaviour and for the steady
states. This agreement deteriorates as $\gamma$ decreases and the boundary of the oscillatory region is approached as can be seen in Fig. \ref{timeseries}b). 
For parameter values in the oscillatory region III 
most simulations (black solid line) die out after a short transient (Fig. \ref{cycles}b)) while the corresponding solutions of the PA SIRS deterministic model (blue solid line) converge to the stable limit cycle for all initial conditions (a typical set is denoted by $B$ in the plot). By choosing initial conditions not far from the stable cycle predicted by the PA SIRS model to avoid extreme susceptible depletion 
during the transient, damped oscillations towards a non trivial equilibrium may also be observed in region III. In Fig. \ref{cycles}a) a plot is shown of one of these surviving simulations (black solid line), together with the solution of the PA equations (blue solid line) for the same parameter values and initial conditions (in the plot denoted by $A$). Thus, instead of an oscillatory phase, the stochastic model on RRGs exhibits in region III a bistability phase, even for large system sizes. 

This failure of the PA model to capture the qualitative behaviour of the simulations on large RRGs is currently being investigated. Extinctions due to finite size are one of the reasons why
the oscillatory phase is seen as an absorbing phase in the stochastic simulations. Indeed, as can be seen in Fig. \ref{cycles}, the oscillations predicted by the PA SIRS deterministic model attain
very small densities of infectives during a significant fraction of the period ($i<10^{-5}$ for initial conditions $B$ in the transient regime). It would be interesting to check whether the more regular oscillations 
that have been observed in the standard PA for some predator-prey models \cite{tome}
persist in stochastic simulations of these models on RRGs. 

However, the breakdown of the PA SIRS model as the boundary of the oscillatory
phase is approached from above and the bistability regime found
in region III show that there are other effects at play.   
The standard pair approximation is only valid for tree-like structures where each node has exactly the same number of contacts and there are no loops, the Bethe lattices. These infinite structures
cannot be simulated on a computer.
On the other hand, classic results of graph theory show that a particular realization of a
RRG-$k$ will contain a large number of loops, of which the large majority are long (with respect to
the average path length), so that locally the graph is essentially tree-like. One would expect 
then the PA to perform well on RRGs, provided they are large enough. 

Increasing system size up to $N=10^7$ we still find suppression of oscillations in region III and significant discrepancies between the transient and steady states of the PA SIRS solutions and the results of the simulations in region II close to the boundary with region III. 
A similar problem of oscillation emergence and suppression and quantitative differences in Monte Carlo simulations versus mean-field approximation and PA of an evolutionary Rock-Scissors-Paper game on different structures was carefully investigated in \cite{szabo}. For this problem, a more accurate multi-site approximation instead of the PA was shown to solve the qualitative and quantitative discrepancies with the simulations. The study of improved models beyond the PA for SIRS dynamics on RRGs will be the subject of future work. 

\subsubsection*{Acknowledgments.}
Financial support from the Foundation of the University of Lisbon 
and the Portuguese Foundation for Science and 
Technology (FCT) under contracts POCI/FIS/55592/2004
and POCTI/ISFL/2/618 is gratefully 
acknowledged. The first author (GR) was also supported by FCT under grant SFRH/BD/32164/2006 and by Calouste Gulbenkian Foundation under its Program 'Stimulus for Research'.

%
%

%
%

%
\end{document}